\documentclass[12pt]{amsart}
\usepackage[alphabetic]{amsrefs}
\usepackage{mathdots, blkarray}
\usepackage{multicol}
\usepackage{graphicx}
\usepackage{amscd}
\usepackage{upgreek}
\usepackage{stmaryrd}
\usepackage{longtable}
\usepackage[T1]{fontenc}
\usepackage{latexsym, amsmath, amssymb, amsthm}
\usepackage[svgnames]{xcolor}
\usepackage{rsfso}
\usepackage[mathscr]{eucal}
\usepackage{mathtools}
\usepackage{mathptmx}
\usepackage{titletoc}
\usepackage{wrapfig}
\usepackage{float}
\usepackage{xypic}
\usepackage{microtype}
\usepackage{dsfont}
\usepackage{xcolor}
\usepackage{color}
\usepackage[colorlinks = true,
            linkcolor  = DarkBlue,
            urlcolor   = DarkRed,
            citecolor  = DarkGreen]{hyperref}
\usepackage{hyperref}
\allowdisplaybreaks	
\usepackage[centering, includeheadfoot, hmargin=1.0in, tmargin=0.5in, 
  bmargin=0.6in, headheight=6pt]{geometry}
\newtheorem{theorem}{Theorem}[section]
\newtheorem{prop}[theorem]{Proposition}

\newtheorem{lem}[theorem]{Lemma}
\newtheorem{coro}[theorem]{Corollary}

\newtheorem{thm}[theorem]{Theorem}

\newtheorem{rem}[theorem]{\rm\textsc{Remark}}
\newtheorem{exam}[theorem]{\rm\textsc{Example}}
\newtheorem{algm}[theorem]{Algorithm}

\newcommand{\ip}[1]{\ensuremath{\left\langle #1 \right\rangle}}

\DeclareMathOperator{\row}{row}

\DeclareMathOperator{\dia}{diag}

\DeclareMathOperator{\Tr}{Tr}
\newcommand{\A}{\mathcal{A}} 
\newcommand{\B}{\mathcal{B}} 
\newcommand{\D}{\mathcal{D}} 
\newcommand{\CC}{\mathcal{C}} 
\newcommand{\C}{\mathbb{C}}

\newcommand{\N}{\mathbb{N}} 
\newcommand{\F}{\mathbb{F}} 
 
\newcommand{\ra}{\longrightarrow}

\newcommand{\hbo}{$\hfill\Diamond$}

\begin{document}
\title{Shape Enumerators of Self-dual NRT Codes over Finite Fields} 
\def\shorttitle{Shape Enumerators of Self-dual NRT Codes over Finite Fields}

\author{Yin Chen}
\address{School of Computer Science \& Technology, Algoma University, Brampton, ON, Canada, L6V 1A3 \& (Current address)
Department of Finance and Management Science, University of Saskatchewan, Saskatoon, SK, Canada, S7N 5A7}
\email{yin.chen@usask.ca}

\author{Runxuan Zhang}
\address{Department of Mathematical and Physical Sciences, Concordia University of Edmonton, Edmonton, AB, Canada, T5B 4E4}
\email{runxuan.zhang@concordia.ab.ca}

\begin{abstract}
We use invariant theory of finite groups to study shape enumerators of self-dual linear codes in
a finite NRT metric space. We provide a new approach that avoids applying Molien's formula to compute all possible shape enumerators. We also explicitly compute the shape enumerators of some low-dimensional self-dual NRT codes over an arbitrary finite field.
\end{abstract}

\date{\today}
\maketitle \baselineskip=15.7pt

\dottedcontents{section}[1.16cm]{}{1.8em}{5pt}
\dottedcontents{subsection}[2.00cm]{}{2.7em}{5pt}

\section{Introduction}
\setcounter{equation}{0}
\renewcommand{\theequation}
{1.\arabic{equation}}
\setcounter{theorem}{0}
\renewcommand{\thetheorem}
{1.\arabic{theorem}}

\noindent Algebraic invariant theory has been a classical subject in modern algebra with a long history, starting with a
faithful representation of a group and exploring all polynomials fixed under the action of the group. 
Invariant theory of finite groups has many substantial ramifications in the study of coding theory, whereas the computational aspects of the invariant rings are indispensable in describing the weight (or shape) enumerators of self-dual linear codes. 
A lovely article \cite{Slo77} published in 1977 explains how to use the technique in invariant theory to study error-correcting self-dual codes over finite fields, inspiring numerous subsequent research work; see, for example, \cite{NRS06} for a general reference of the theory of self-dual codes and invariant theory. 
 

To capitalize on invariant theory in coding theory, a MacWilliams' theorem (also known as a MacWilliams' identity) usually serves as the first key step. This theorem proves that the weight (or shape) enumerators of the dual of a code can be expressed by the weight (or shape) enumerators of the code. Consider an $r$-dimensional code $\CC$ in the $n$-dimensional Hamming metric space over a finite field $\F_q$ and denote by $W_\CC(x,y)$ the weight enumerator of $\CC$. The original MacWilliams' theorem states that
\begin{equation}
\label{ }
W_{\CC^\top}(x,y)=\frac{1}{q^r}\cdot W_\CC(x+(q-1)y,x-y)
\end{equation} 
where $\CC^\top$ denotes the dual code of $\CC$; see \cite{Mac63} or \cite[Theorem 4]{Slo77}.
Using this identity and the classical Molien's formula in invariant theory, Gleason \cite{Gle71} completely described the weight enumerators of all the binary doubly-even self-dual codes, showing that these enumerators can be expressed algebraically by two polynomials fixed by the action of a finite group.  

Recently, a non-Hamming metric, called the \textit{NRT metric}, on the vector space of matrices over a finite field was introduced by
\cite{Nie91} and \cite{RT97}, stemming from the maximization problem in finite vector space and generalizing Reed-Solomon codes in finite matrix vector spaces. Several coding-theoretic subtopics within this metric have been studied extensively; see \cite{Alv11, BF12, BP15, DS02, DS04}, and \cite{PB10}.  
In particular, a MacWilliams theorem for the linear codes in an NRT space was proved in \cite[Theorem 3.1]{DS02}.
Based on this theorem and using the classical Molien's theorem in invariant theory, Santos and Alves have described the shape enumerators of binary self-dual NRT codes (i.e., the ground field is $\F_2$) in \cite{SA20}.

The present article continues to explore the shape enumerators of self-dual NRT codes over an arbitrary finite field $\F_q$.
The MacWilliams' theorem for NRT codes mentioned above implies that the shape enumerator of a self-dual NRT code over $\F_q$ must be an invariant polynomial of a group $G=\ip{\upsigma}$ of order $2$ with a high-dimensional faithful complex representation $V$. This indicates that describing the shape enumerators of all self-dual NRT codes over $\F_q$ could be transformed to the question of finding all invariant polynomials of $G$ with the representation $V$.

We provide a new approach, via representation theory and invariant theory of the cyclic group $C_2$, to give a minimal generating set for the invariant ring $\C[V]^G$, avoiding the use of Molien's theorem. Roughly speaking, our first step is to find an
equivalent $G$-representation $W$ for $V$ so that the invariant ring $\C[W]^G$ is relatively easier to compute; the second step is devoted to finding a homogeneous generating set $\A$ for $\C[W]^G$;  the third step uses some suitable linear maps to transfer elements of $\A$ to a homogeneous generating set $\B$ for $\C[V]^G$. Note that $\C[W]^G\neq \C[V]^G$ as $\C$-algebras but $\C[W]^G$ is isomorphic to $\C[V]^G$ and the shape enumerator  of any self-dual NRT code over $\F_q$ can be algebraically expressed  by
elements of $\B$.

Throughout this article we are working within the NRT space $M_{n,m}(\F_q)$ and assume that $m$ is even.  Section \ref{sec2} below presents some fundamental definitions and facts about NRT codes, including NRT weight, NRT metric, NRT spaces, self-dual NRT codes, the shape enumerator of an NRT code, and the MacWilliams Theorem (see (\ref{Mac})).  
A key fact we observe shows that the trace of the generator $\upsigma$ of the group $G$ is $1$; see Lemma \ref{lem2.1}.
In Section \ref{sec3}, we first determine the equivalent $G$-representation $W$ of the standard representation $V$, on which 
the resulting matrix $\upsigma_W$ of $\upsigma$ is the diagonal matrix along with the first $\frac{m}{2}+1$ entries being 1 and  
the last $\frac{m}{2}$ entries being $-1$. This fact allows us to find a very quick way to compute $\C[W]^G$, exhibiting a simple homogeneous generating set $\A$ for $\C[W]^G$; see Theorem \ref{thm3.2}.  We also provide an algorithm to transfer the set $\A$
to a homogeneous generating set $\B$ for $\C[V]^G$; see Algorithm \ref{algm}. Section \ref{sec4} consists of  four examples. The first three examples illustrate Algorithm \ref{algm} and Corollary \ref{main} for $m=2, 4$, and $6$. The last example articulates the "normal subgroup" technique from invariant theory to find the shape enumerators of some even binary self-dual codes that have been obtained in \cite{SA20} via Molien's theorem.

\section{NRT Codes over Finite Fields}\label{sec2}
\setcounter{equation}{0}
\renewcommand{\theequation}
{2.\arabic{equation}}
\setcounter{theorem}{0}
\renewcommand{\thetheorem}
{2.\arabic{theorem}}

\subsection{Finite NRT spaces} Let $\F_q$ be a finite field of order $q$ and $M_{n,m}(\F_q)$ be the vector space of all $n\times m$ matrices over $\F_q$, where $n,m\in\N^+$. The \textit{weight} of a row vector $v=(v_1,v_2,\dots,v_m)\in M_{1,m}(\F_q)$ is defined as
\begin{equation}
\label{ }
\uprho(v):=\max\{j\in\{1,2,\dots,m\}\mid v_j\neq 0\}
\end{equation}
and we define  the \textit{weight} of a matrix $A\in M_{n,m}(\F_q)$ to be
\begin{equation}
\label{ }
\uprho(A):=\sum_{i=1}^n\uprho(\row_i(A)).
\end{equation}
The \textit{NRT metric} on $M_{n,m}(\F_q)$ is defined by
$$d_\uprho(A,B):=\uprho(A-B)$$
for all $A,B\in M_{n,m}(\F_q)$. We observe that the NRT metric is a non-Hamming metric unless $m=1$ for which the NRT metric coincides with the classical Hamming metric. 

Given two matrices $A=(a_{ij}), B=(b_{ij})\in M_{n,m}(\F_q)$, we may define an inner product on $M_{n,m}(\F_q)$ endowed with  the NRT metric by
\begin{equation}
\label{ }
\ip{A,B}:=\sum_{i=1}^n \ip{\row_i(A), \row_i(B)}
\end{equation}
where $\ip{\row_i(A), \row_i(B)}:=\sum_{j=1}^m a_{ij}\cdot b_{i,m-j+1}$. Together with this inner product and the NRT metric, the vector space  $M_{n,m}(\F_q)$ is called an \textit{NRT space} over $\F_q$. 
Clearly, $\ip{\row_i(A), \row_i(B)}=\ip{\row_i(B), \row_i(A)}$ for all $i=1,2,\dots,n$, thus $\ip{A,B}=\ip{B,A}$. See, for example, \cite{RT97, BGL95} or \cite[Section 2]{SA20} for more details.

\subsection{Finite NRT codes}

An $r$-dimensional vector subspace $\CC$ of $M_{n,m}(\F_q)$ is called an \textit{NRT  code} over $\F_q$, and its dual code 
is defined as
\begin{equation}
\label{ }
\CC^\top:=\{A\in M_{n,m}(\F_q)\mid \ip{A,B}=0,\textrm{ for all }B\in\CC\}
\end{equation}
which is a vector subspace of $M_{n,m}(\F_q)$ with dimension $\dim(M_{n,m}(\F_q))-\dim(\CC)=mn-r$.

We say that $\CC$ is \textit{self-dual} if $\CC=\CC^\top.$ Clearly, if $\CC$ is a self-dual code in
$M_{n,m}(\F_q)$, then $mn$ is divisible by $2$ and $\dim(\CC)=\frac{mn}{2}$.

Let $A\in M_{n,m}(\F_q)$ be a matrix. The \textit{shape} $e_A$ of $A$ is an integral vector defined as
$$e_A:=(e_1,e_2,\dots,e_m)\in\N^m$$
where $e_j$ denotes the cardinality of the set $\{i\in\{1,2,\dots,n\}\mid \uprho(\row_i(A))=j\}$.  Note that
\begin{equation}
\label{ }
\uprho(A)=\sum_{j=1}^m j\cdot e_j.
\end{equation}
We also define $|e_A|:=\sum\limits_{j=1}^m e_j$ and $e_0:=n-|e_A|$. Thus, a shape vector
$(e_1,e_2,\dots,e_m)$ induces a partition of $n$ into a sum of $m+1$ parts. All such partitions can be denoted by
\begin{equation}
\label{ }
\Delta_{m,n}:=\{(e_1,e_2,\dots,e_m)\in\N^m\mid e_0+e_1+\dots+e_m=n\}.
\end{equation}
More details about NRT linear codes can be found in \cite{BP15} and \cite{DS02}.

\subsection{Shape enumerators}
Let $\CC$ be an NRT code in $M_{n,m}(\F_q)$. The \textit{shape enumerator} of $\CC$ is a homogeneous
complex polynomial of degree $n$ in $m+1$ variables defined by
\begin{equation}
\label{ }
H_\CC(x):=\sum_{(e_1,\dots,e_m)\in \Delta_{m,n}} c_e\cdot x_0^{e_0}x_1^{e_1}\cdots x_m^{e_m}
\end{equation}
where $c_e$ denotes the cardinality of the set $\{A\in\CC\mid e_A=(e_1,\dots,e_m)\}$.

The MacWilliams' identity for finite NRT codes was proved in \cite[Theorem 3.1]{DS02} (or see \cite{BP15}), establishing a close connection between the shape enumerators of an NRT code $\CC$ and its dual code $\CC^\top$:
\begin{equation}
\label{Mac}
H_{\CC^\top}(x)=\frac{1}{|\CC|}\cdot H_\CC(g\cdot x)
\end{equation}
where $x=(x_0,x_1,\dots,x_m)$ and $g:=(g_{st})$ denotes the invertible real matrix of size $m+1$ with $0\leqslant s,t\leqslant m$, defined by
\begin{equation}
\label{eq2.9}
g_{st}:=\begin{cases}
     1, & t=0, \\
     q^{t-1}(q-1), & 0<t\leqslant m-s,\\
     -q^{t-1},&t+s=m+1,\\
     0,&t+s>m+1.
\end{cases}
\end{equation}
Note that $g^2=q^m\cdot I_{m+1}$ for all $m,n\in\N^+$, thus
\begin{equation}
\label{ }
\left(q^{-\frac{m}{2}}\cdot g\right)^2=I_{m+1}
\end{equation}
where $I_{m+1}$ denotes the identity matrix of size $m+1$.

Suppose that $\CC$ is a self-dual code. Then $\dim(\CC)=\frac{nm}{2}$, and so either $n$
or $m$ must be even. Throughout the rest of the paper, we assume that $m$ is even. Thus $|\CC|=q^{\frac{nm}{2}}=\left(q^{\frac{m}{2}}\right)^n$. Hence,
\begin{equation}
\label{ }
H_{\CC}(x)=H_\CC\left(q^{-\frac{m}{2}}\cdot g\cdot x\right).
\end{equation}
This means that the shape enumerator $H_{\CC}(x)$  is an invariant polynomial
of the cyclic group of order $2$ generated by $q^{-\frac{m}{2}}\cdot g$. 

\begin{lem}\label{lem2.1}
The trace of $q^{-\frac{m}{2}}\cdot g$ is $1$ when $m$ is even. 
\end{lem}

\begin{proof}
Assume that $m=2u$ for some $u\in\N^+$. Note that $g_{ii}=0$ for all $i\geqslant u+1$. Thus
\begin{eqnarray*}
\Tr(g)&=&g_{00}+g_{11}+\cdots+g_{uu}+g_{u+1,u+1}+\dots+g_{mm}\\
&=&1+(q-1)+q(q-1)+\cdots+q^{u-1}(q-1)\\
&=&q^u=q^{\frac{m}{2}}.
\end{eqnarray*}
Hence, $\Tr(q^{-\frac{m}{2}}\cdot g)=q^{-\frac{m}{2}}\cdot\Tr(g)=q^{-\frac{m}{2}}\cdot q^{\frac{m}{2}}=1.$
\end{proof}

\section{Representations and Polynomial Invariants of $C_2$} \label{sec3}
\setcounter{equation}{0}
\renewcommand{\theequation}
{3.\arabic{equation}}
\setcounter{theorem}{0}
\renewcommand{\thetheorem}
{3.\arabic{theorem}}

\subsection{Representation theory of $C_2$}
Let $C_2=\ip{\upsigma}$ be the cyclic group of order $2$. By representation theory, we see that there only exists two irreducible complex representations of $C_2$: the trivial representation $V_1$ and the sign representation $V_{2}$, which both are 1-dimensional and defined by
\begin{equation}
\label{ }
\upsigma\mapsto 1\textrm{ and }\upsigma\mapsto -1
\end{equation}
respectively.

Suppose that $W$ denotes a finite-dimensional complex representation of $C_2$. Then there exist two non-negative integers 
$a,b\in\N$ such that
\begin{equation}
\label{ }
W\cong a\cdot V_1\oplus b\cdot V_2.
\end{equation}
In other words, after choosing a suitable basis for $W$, the resulting matrix of $\upsigma$ on $W$ is 
\begin{equation}
\label{ }
\upsigma_W=\dia\{1,\dots,1,-1,\dots,-1\}
\end{equation}
with $a$ copies of $1$ and $b$ copies of $-1$. Moreover, if $W$ is faithful, then $b\geqslant 1$.

Now we let $G$ be the cyclic group of order $2$ generated by $q^{-\frac{m}{2}}\cdot g$ and
$V$ be the standard complex representation of $G$. Then $V$ is faithful, and so we may write 
\begin{equation}
\label{eq3.4}
V\cong a\cdot V_1\oplus b\cdot V_2
\end{equation}
for some $a\in\N$ and $b\in\N^+$. Note that $\dim(V)=m+1$, thus
\begin{equation}
\label{eq3.5}
a+b=m+1.
\end{equation}
By Lemma \ref{lem2.1}, we have
\begin{equation}
\label{eq3.6}
a-b=1.
\end{equation}
Combining these two equations (\ref{eq3.5}) and (\ref{eq3.6}) yields
\begin{equation}
\label{eq3.7}
a=\frac{m}{2}+1\textrm{ and  }b=\frac{m}{2}.
\end{equation}

\subsection{Polynomial invariants of $C_2$} 
Let $k$ be a field of any characteristic  and $W$ be a finite-dimensional faithful  representation of a finite group $G$ over $k$. The dual representation $W^*$ induces an algebraic action of $G$ on the symmetric algebra $k[W]$ on $W^*$. Choosing a basis $\{x_1,x_2,\dots,x_n\}$ for $W^*$, we may identify $k[W]$ with the polynomial ring $k[x_1,x_2,\dots,x_n]$. The subring of $k[W]$ consisting of all polynomials fixed by the action of $G$ is called the \textit{invariant ring} of $G$ on $W$, and denoted by $k[W]^G$. More precisely,
\begin{equation}
\label{ }
k[W]^G=k[x_1,x_2,\dots,x_n]^G:=\{f\in k[x_1,x_2,\dots,x_n]\mid g(f)=f,\textrm{ for all }g\in G\}
\end{equation}
which is the main object of study in polynomial invariant theory of finite groups. A theorem due to Emmy Noether in 1923 states that
$k[W]^G$ is a finitely generated $\N$-graded commutative algebra over $k$; see \cite[Proposition 3.0.1]{DK15} for a modern proof of this theorem. 

To understand the algebraic structure of an invariant ring $k[W]^G$,  the study of two fundamental questions plays a core role 
 in invariant theory. The first question is about how to find a minimal generating set for $k[W]^G$, and the second one asks how to find a set of generating relations among these generators. The invariant ring $k[W]^G$ is called \textit{modular} if the characteristic 
 of $k$ divides the order of $G$; otherwise, \textit{nonmodular}.  The nonmodular case includes two subcases: (1)  the characteristic of $k$ is zero; (2) the characteristic of $k$ is positive but doesn't divide the order of $G$. Nonmodular invariant theory has been understood very well while modular invariant theory is a challenging topic; see for example, \cite{DK15} or \cite{CW11} for general references of invariant theory of finite groups. 

The following general result is crucial for the present article and it might be well-known to experts in invariant theory but we didn't find a suitable reference including a proof. We will provide a constructive proof later in Subsection \ref{subsec3.3}.

\begin{prop}\label{prop3.1}
Let $W_1$ and $W_2$ be two equivalent $n$-dimensional representations of a finite group $G$ over a field $k$. Then $k[W_1]^G$ and $k[W_2]^G$ are isomorphic as $\N$-graded algebras over $k$.
\end{prop}

Throughout this article, if $A=(a_{ij})_{n\times n}$ is a group element, we assume that the action of $A$ on a polynomial $f(x_1,x_2,\dots,x_n)$ is given by
\begin{eqnarray*}
A(x_1) & = & a_{11}x_1+a_{21}x_2+\cdots+a_{n1}x_n \\
A(x_2) & = & a_{12}x_1+a_{22}x_2+\cdots+a_{n2}x_n \\ 
\vdots\quad&\vdots&\quad\quad\quad\quad\vdots\\
A(x_n) & = & a_{1n}x_1+a_{2n}x_2+\cdots+a_{nn}x_n.
\end{eqnarray*}

In our situation, recall that the ground field $k=\C$ and the group $G$ is the cyclic group of order $2$ generated by 
\begin{equation}
\label{eq3.9}
\upsigma:=q^{-\frac{m}{2}}\cdot g.
\end{equation}
We have seen from (\ref{eq3.7}) that
\begin{equation}
\label{ }
V\cong W:=\left(\frac{m}{2}+1\right)\cdot V_1\oplus \frac{m}{2}\cdot V_2.
\end{equation}
Thus  as commutative $\N$-graded finitely generated $\C$-algebras, we have
\begin{equation}
\label{eq3.11}
\C[V]^G\cong \C[W]^G
\end{equation}
are isomorphic, by Proposition \ref{prop3.1}.

To compute $\C[V]^G$, we first compute $\C[W]^G$ and then we provide an algorithm in next subsection to transfer 
a generating set of $\C[W]^G$ to a generating set for $\C[V]^G$. 

We may assume that $\C[W]=\C[x_0,x_1,\dots,x_{m}]$, and note that $m$ is an even number. 

\begin{thm}\label{thm3.2}
$\C[W]^G=\C[x_0,x_1,\dots,x_{m}]^G=\C[x_0,x_1,\dots,x_{\frac{m}{2}}, z_{ij}:=x_ix_j\mid \frac{m}{2}+1\leqslant i\leqslant j\leqslant m]$.
\end{thm}

\begin{proof}
First of all, Noether's bound theorem tells us that $\C[W]^G$ can be generated by finitely many homogeneous invariant polynomials of degree at most $|G|=2$; see \cite{Fle00} or \cite[Theorem 3.5.1]{CW11}. As the resulting matrix $\upsigma_W$ of $\upsigma$ on $W$ is a diagonal matrix along with the first $\frac{m}{2}+1$ entries 1 and the remaining $\frac{m}{2}$ entries $-1$, we see that
$$x_0,x_1,\dots,x_{\frac{m}{2}}$$
are invariant and furthermore, they are clearly indecomposable; namely, these invariants  can not written as 
the product of two invariants of lower degrees, because their degrees are 1. 

Suppose that $\ell=a_1\cdot x_{\frac{m}{2}+1}+\cdots+a_{\frac{m}{2}}\cdot x_m$ denotes any linear invariant where all $a_i$ are complex numbers. Then $\ell=\sigma_W(\ell)=a_1\cdot (-x_{\frac{m}{2}+1})+\cdots+a_{\frac{m}{2}}\cdot (-x_m)=-\ell$. Thus
$2\ell=0$ and $\ell=0$. In other words, there are no nonzero linear invariants involving any one of 
$\{x_i\mid \frac{m}{2}+2\leqslant i\leqslant m\}$.

Hence, we only need to consider the quadratic invariants in $\{x_i\mid \frac{m}{2}+1\leqslant i\leqslant m\}$. We define
\begin{equation}
\label{ }
z_{ij}:=x_ix_j
\end{equation}
for $\frac{m}{2}+1\leqslant i\leqslant j\leqslant m$. Since $\upsigma_W(z_{ij})=(-x_i)(-x_j)=x_ix_j=z_{ij}$ and $\upsigma_W(x_i)=-x_i\neq x_i$ for $\frac{m}{2}+1\leqslant i\leqslant j\leqslant m$, we see that all $z_{ij}$ are indecomposable invariants. 
Therefore, $\C[W]^G$ can be generated minimally by $x_0,x_1,\dots,x_{\frac{m}{2}}$ and $z_{ij}$, where $\frac{m}{2}+1\leqslant i\leqslant j\leqslant m$.
\end{proof}

\begin{coro}
If $m$ is an even integer greater than or equal to $4$, then $\C[V]^G\cong \C[W]^G$ is not a polynomial algebra.  
\end{coro}

\begin{proof}
For an even integer $m\geqslant 4$, Theorem \ref{thm3.2} shows that the set $$\left\{x_0,x_1,\dots,x_{\frac{m}{2}}, z_{ij}\mid \frac{m}{2}+1\leqslant i\leqslant j\leqslant m\right\}$$ is a minimal generating set of $\C[W]^G$, which has cardinality 
$\begin{pmatrix}
\frac{m}{2}+1\\
      2  
\end{pmatrix}$. This number is strictly greater than the Krull dimension $m+1$ of $\C[W]^G$ for all even $m\geqslant 4$. Thus, $\C[W]^G$ is not a polynomial algebra. By (\ref{eq3.11}) (or Proposition \ref{prop3.1}), we see that $\C[V]^G$ and $\C[W]^G$
are isomorphic as $\C$-algebras. Hence, $\C[V]^G\cong \C[W]^G$ is also not a polynomial algebra.
\end{proof}

\subsection{An algorithm} \label{subsec3.3}

Let's begin with a constructive proof of Proposition \ref{prop3.1}.

\begin{proof}[Proof of Proposition \ref{prop3.1}]
As $W_1$ and $W_2$ are equivalent, there exists an invertible matrix $T\in GL_n(k)$ such that 
\begin{equation}
\label{ }
g_{W_2}=T^{-1}\circ g_{W_1}\circ T
\end{equation}
for all $g\in G$. Thus
\begin{equation}
\label{ }
(g_{W_2}^{-1})^t=T^t\circ (g_{W_1}^{-1})^t\circ (T^{-1})^t=T^t\circ (g_{W_1}^{-1})^t\circ (T^{t})^{-1}
\end{equation}
which means that the two dual representations $W_1^*$ and $W_2^*$ are also equivalent as $G$-representations. 

Let $T_d$ be the induced invertible matrix of $T$ on the $d$-th symmetric power $k[x_1,\dots,x_n]_d$.
For instance, we have seen that $T_1=(T^{t})^{-1}$. Then
\begin{equation}
\label{ }
g_{S_2^d}=T_d^{-1}\circ g_{S_{1}^d}\circ T_d
\end{equation}
for all $g\in G$, where $S^d_1$ and $S^d_2$ denote the $d$-th symmetric power representations induced by $W_1$ and $W_2$ respectively. Thus $S^d_1$ and $S^d_2$ are equivalent $G$-representations. 

For each $d\in\N^+$, if $f\in k[W_1]_d^G$, then 
$$g_{S_2^d}(T_d^{-1}(f))=\left(g_{S_2^d}\circ T_d^{-1}\right)(f)=\left(T_d^{-1}\circ g_{S_{1}^d}\right)(f)=T_d^{-1}(f).$$
Thus, the map $f\mapsto T_d^{-1}(f)$ gives rise to a linear isomorphism from $k[W_1]_d^G$ to $k[W_2]_d^G$, where
the action $T_d^{-1}(f)$ is defined by 
\begin{equation}
\label{ }
T_d^{-1}(f(x_1,\dots,x_n)):=f\left(T_1^{-1}(x_1),T_1^{-1}(x_2),\dots,T_1^{-1}(x_n)\right)
\end{equation}
for all $f(x_1,x_2,\dots,x_n)\in k[W_1]_d^G$. Clearly, this isomorphism can be extended to a $k$-algebra isomorphism from $k[W_1]^G$ to $k[W_2]^G$.
\end{proof}

\begin{rem}{\rm
Here we would like to record another proof of Proposition \ref{prop3.1} due to the anonymous referee. 
A linear isomorphism $\varphi:W_1\ra W_2$ gives us a linear isomorphism $\varphi^*:W_2^*\ra W_1^*$, which induces an algebra isomorphism $\varphi^s:k[W_2]\ra k[W_1]$, by the universal property of the symmetric algebra. The isomorphism $\varphi^s$
is given on homogeneous generators of $k[W_2]$ by $y_{i_1}\cdots y_{i_d}\mapsto \varphi^*(y_{i_1})\cdots \varphi^*(y_{i_d})$.
As $\varphi$ commutes with the actions of $G$, it follows that both $\varphi^*$ and $\varphi^s$ commute with the actions of $G$.
Hence, the isomorphism $\varphi^s$ restricts to an isomorphism between $k[W_2]^G$ and $k[W_1]^G$.
\hbo}\end{rem}

The above proof of Proposition \ref{prop3.1} actually provides us an algorithm that can transfer a generating set of $k[W_1]^G$ into a generating set for $k[W_2]^G$ by applying some $T_d$ where $d$ denotes the Noether bound of $k[W_1]^G$; in fact, $d=|G|$ for the non-modular case (see \cite{Fle00}) and $d=\dim(W_1)(|G|-1)$ for the modular case; see  \cite{Sym11}.

Suppose that $W_1$ and $W_2$ are two equivalent  representations of a finite group $G$ over a field $k$ and
$\A$ denotes a homogeneous generating set of $k[W_1]^G$. Assume that $d$ denotes the maximal degree of elements of $\A$ and arrange the elements of $\A$ in the ascending order of degree, i.e., $\A=\{f_1,f_2,\dots,f_s\}$ with $\deg(f_1)\leqslant \deg(f_2)\leqslant \cdots\leqslant \deg(f_s)=d$.

\begin{algm} \label{algm}
We may construct a homogeneous generating set $\B$ for $k[W_2]^G$ from $\A$ by performing the following steps:
\begin{enumerate}
  \item Find $T\in GL_n(k)$ such that $g_{W_2}=T^{-1}\circ g_{W_1}\circ T$
for all $g\in G$;
  \item Let $\B:=\emptyset, f:=f_1$ and repeat Steps $(2) - (3)$ where $f$ runs over $\A$;
  \item Compute the invertible matrix $T_{\deg(f)}^{-1}$ and add the image $T_{\deg(f)}^{-1}\cdot f$ into $\B$;
  \item After $m$ steps, this algorithm terminates and $\B$ is a homogeneous generating set of $k[W_2]^G$.
\end{enumerate}
\end{algm}
\begin{proof}
The correctness of this algorithm is immediate from the proof of Proposition \ref{prop3.1} . 
\end{proof}

\begin{rem}{\rm
The complexity of this algorithm mainly depends on the dimension $n$ of the representation space $W_1$ (or $W_2$) and the cardinality $|\A|$ of the generating set $\A$. Usually when $n$ increases, finding the 
invertible matrix $T$ in the first step would cost more time, as well as the cardinality $|\A|$ might become very large. Actually,
if the invariant ring $k[W_1]^G$ is not a polynomial algebra, it may contain many indecomposable invariants and finding a minimal generating set for this invariant ring could be extremely challenging, even though for the nonmodular case.
\hbo}\end{rem}

\begin{coro}\label{main}
Let $\upsigma$ be the matrix defined in $(\ref{eq3.9})$ and $m$ be even. 
\begin{enumerate}
  \item There exists an invertible matrix $T$ of size $m+1$ such that $$\upsigma=T^{-1}\cdot \dia\{1,\dots,1,-1,\dots,-1\}\cdot T,$$ with $\frac{m}{2}+1$ copies of $1$ and $\frac{m}{2}$ copies of $-1$.
  \item The shape enumerator of each self-dual NRT code in $M_{n,m}(\F_q)$ is a polynomial in the images of 
  $x_0,x_1,\dots,x_{\frac{m}{2}}, x_ix_j$ with $\frac{m}{2}+1\leqslant i\leqslant j\leqslant m$,
  under the linear transformation $T^t$.
\end{enumerate} 
\end{coro}

\section{Examples} \label{sec4}
\setcounter{equation}{0}
\renewcommand{\theequation}
{4.\arabic{equation}}
\setcounter{theorem}{0}
\renewcommand{\thetheorem}
{4.\arabic{theorem}}

\noindent This section contains four examples for which the first three examples illustrate Algorithm \ref{algm} and Corollary \ref{main} and the last one explains how to use the "normal subgroup" technique from invariant theory (see \cite[Section 1.10]{CW11}), instead of using Molien's formula, to compute the shape enumerators of some special NRT self-dual codes.

\begin{exam}{\rm Consider any $n\in\N^+$ and $m=2$. In this case, it follows from (\ref{eq2.9}) that
\begin{equation}
\label{ }
\sigma_W=\dia\{1,1,-1\}\textrm{ and } \sigma_{V}=\frac{1}{q}\begin{pmatrix}
  1    & q-1  &q(q-1) \\
    1  & q-1 &-q\\
    1&-1&0
\end{pmatrix}
\end{equation}
Theorem \ref{thm3.2} shows that
$$\C[W]^G=\C[x_0,x_1,x_{2}]^G=\C[x_0,x_1,x_2^2]$$
is a polynomial algebra. Note that
\begin{equation}
\label{ }
\sigma_{V}=\begin{pmatrix}
  \frac{1}{q}    & \frac{q-1}{q}  &q-1 \\
    \frac{1}{q}  & \frac{q-1}{q} &-1\\
    \frac{1}{q}&-\frac{1}{q}&0
\end{pmatrix}=\begin{pmatrix}
   \frac{q+1}{q}   & \frac{q-1}{q} & q-1 \\
     1 & -1 &q\\
    1  &-1 &-q
\end{pmatrix}^{-1}\begin{pmatrix}
    1  &0&0    \\
     0 &1&0\\
     0&0&-1  
\end{pmatrix}\begin{pmatrix}
   \frac{q+1}{q}   & \frac{q-1}{q} & q-1 \\
     1 & -1 &q\\
    1  &-1 &-q
\end{pmatrix}.
\end{equation}
Thus, the invertible matrices $T$ and $T_1^{-1}$ we are looking for are:
$$T=\begin{pmatrix}
   \frac{q+1}{q}   & \frac{q-1}{q} & q-1 \\
     1 & -1 &q\\
    1  &-1 &-q
\end{pmatrix}\textrm{ and }T_1^{-1}=T^t=\begin{pmatrix}
   \frac{q+1}{q} &1&1\\
   \frac{q-1}{q} &-1&-1\\
   q-1&q&-q\\
\end{pmatrix}.$$
By Algorithm \ref{algm}, we obtain a minimal generating set $\B=\{g_1,g_2,g_3\}$ for $\C[V]^G$, where
\begin{eqnarray}
g_1 &:= &T_1^{-1}(x_0)= \frac{q+1}{q}\cdot x_0+ \frac{q-1}{q}\cdot x_1+(q-1)\cdot x_2 \nonumber\\
g_2 &:= &T_1^{-1}(x_1)= x_0-x_1+q\cdot x_2 \\
g_3&:=& T_2^{-1}(x_2^2)=\left(T_1^{-1}(x_2)\right)^2= (x_0-x_1-q\cdot x_2)^2.\nonumber
\end{eqnarray}
Note that the invariant ring $\C[V]^G$ may have  a simpler generating set. This means that
our generating set $\B$ may be not simplest but it is the most direct. 

Suppose $q=2$. Let's compare our generating set $\B=\{g_1,g_2,g_3\}$ with the generating set 
$\D:=\{\varphi_1,\varphi_2,\varphi_3\}$ already obtained in \cite[Theorem 17]{SA20}. In this special case, 
\begin{eqnarray*}
 \varphi_1=x_0+x_2,&\varphi_2=x_0+x_1,&\varphi_3=x_0^2+x_1^2+2x_2^2.
\end{eqnarray*}
A direct computation shows that
\begin{eqnarray*}
g_1&=&\varphi_1+\frac{1}{2}\varphi_2\\
g_2&=&2\varphi_1-\varphi_2\\
g_3&=&4\varphi_3-3\varphi_2^2+4\varphi_1\varphi_2-4\varphi_1^2.
\end{eqnarray*}
Conversely, 
\begin{eqnarray*}
\varphi_1 & = & \frac{1}{2}g_1+\frac{1}{4}g_2 \\
\varphi_2& = & g_1-\frac{1}{2}g_2\\
\varphi_3& = &\frac{1}{2}g_1^2-\frac{1}{2}g_1g_2+\frac{3}{8}g_2^2+\frac{1}{4}g_3. 
\end{eqnarray*}
This reproves that the two sets $\B$ and $\D$ both are generating sets for the invariant ring $\C[V]^G$. However, the difference between our constructing method and the way appeared in \cite{SA20} is that our generating set $\B$ works for all $q$.
\hbo}\end{exam}

\begin{exam}{\rm
Consider any $n\in\N^+$ and $m=4$. By (\ref{eq2.9}), we see that
\begin{equation}
\label{ }
\sigma_W=\dia\{1,1,1,-1,-1\}\textrm{ and } \sigma_{V}=\frac{1}{q^2}\begin{pmatrix}
  1    & q-1  &q(q-1) &q^2(q-1)&q^3(q-1)\\
    1  & q-1 &q(q-1)&q^2(q-1)&-q^3\\
    1&q-1&q(q-1)&-q^2&0\\
    1&q-1&-q&0&0\\
    1&-1&0&0&0
\end{pmatrix}
\end{equation}
Theorem \ref{thm3.2} implies that 
$$\C[W]^G=\C[x_0,x_1,x_{2},x_3,x_4]^G=\C[x_0,x_1,x_2,x_3^2,x_4^2,x_3x_4]$$
is not a polynomial algebra but a hypersurface, subject to the unique relation
\begin{equation}
\label{ }
(x_3x_4)^2=(x_3^2)(x_4^2).
\end{equation}
Let's define
\begin{equation}
\label{ }
T:=\begin{pmatrix}
(q^2 + 1)/q^2 &   (q - 1)/q^2  &    (q - 1)/q     &     q - 1    &    q^2 - q    \\
1    &      q - 1  &    2q^2 - q    &       -q^2        &      0\\
 1    &         -1    &          0        &      0     &       q^2\\
 (-q^2 + 1)/q^2  &  (q - 1)/q^2   &   (q - 1)/q    &      q - 1    &    q^2 - q\\
 1      &    q - 1     &        -q     &      -q^2       &       0
\end{pmatrix}
\end{equation}
and a direct computation shows that
$$\sigma_{V}=T^{-1}\cdot \sigma_{W}\cdot T$$
Hence, the invertible matrix $T_1^{-1}=T^t$ and $\C[V]^G$ can be generated minimally by
\begin{eqnarray*}
g_1 & := & T_1^{-1}(x_0)= \frac{q^2 + 1}{q^2}\cdot x_0 +   \frac{q - 1}{q^2}\cdot x_1  +  \frac{q-1}{q} \cdot x_2     +     (q - 1)\cdot x_3    +    (q^2 - q)\cdot x_4\\
g_2 & := & T_1^{-1}(x_1)= x_0+(q-1)\cdot x_1+(2q^2 - q)\cdot x_2-q^2\cdot x_3\\
g_3&:=&T_1^{-1}(x_2)=x_0-x_1+q^2\cdot x_4\\
g_4&:=& [T_1^{-1}(x_3)]^2=\left(-\frac{q^2 + 1}{q^2}\cdot x_0 +   \frac{q - 1}{q^2}\cdot x_1  +  \frac{q-1}{q} \cdot x_2     +     (q - 1)\cdot x_3    +    (q^2 - q)\cdot x_4\right)^2\\
g_5&:=& [T_1^{-1}(x_4)]^2= (x_0+(q-1)\cdot x_1-q\cdot x_2-q^2\cdot x_3)^2\\
g_6&:=& T_1^{-1}(x_3)\cdot T_1^{-1}(x_4)=
\end{eqnarray*}\vspace{-7mm}
{\small$$\left(-\frac{q^2 + 1}{q^2}\cdot x_0 +   \frac{q - 1}{q^2}\cdot x_1  +  \frac{q-1}{q} \cdot x_2     +     (q - 1)\cdot x_3    +    (q^2 - q)\cdot x_4\right)\cdot \Big(x_0+(q-1)\cdot x_1-q\cdot x_2-q^2\cdot x_3\Big)$$}
with the unique relation $g_6^2=g_4\cdot g_5.$
\hbo}\end{exam}

\begin{exam}{\rm
Let $n\in\N^+$ and $m=6$. It follows from (\ref{eq2.9}) that $\sigma_W=\dia\{1,1,1,1,-1-1,-1\}$ and 
\begin{equation}
\label{ }
\sigma_{V}=\frac{1}{q^3}\begin{pmatrix}
  1    & q-1  &q(q-1) &q^2(q-1)&q^3(q-1)&q^4(q-1)&q^5(q-1)\\
    1  & q-1 &q(q-1)&q^2(q-1)&q^3(q-1)&q^4(q-1)&-q^5\\
    1&q-1&q(q-1)&q^2(q-1)&q^3(q-1)&-q^4&0\\
    1&q-1&q(q-1)&q^2(q-1)&-q^3&0&0\\
    1&q-1&q(q-1)&-q^2&0&0&0\\
    1&q-1&-q&0&0&0&0\\
    1&-1&0&0&0&0&0
\end{pmatrix}.
\end{equation}
By Theorem \ref{thm3.2}, we see that 
$$\C[W]^G=\C[x_0,x_1,\dots,x_6]^G=\C[x_0,x_1,x_2,x_3,x_4^2,x_5^2,x_6^2,x_4x_5,x_4x_6,x_5x_6]$$
is a complete intersection (i.e., the number of generators minus the number of generating relations is equal to the Krull dimension of the invariant ring), subject to the three generating relations: $$(x_ix_j)^2=x_i^2\cdot x_j^2$$ for $4\leqslant i<j\leqslant 6$.
We define
$$
T:=\begin{pmatrix}
1  & q - 1 &  q^3 + q^2 - q &  q^3 - q^2 &  q^4 - q^3 &  -q^4 &  0\\
1  & -1 &  0 &  0 &  0 &  0 &  q^3\\
(q^3 + 1)/q^3 &  (q - 1)/q^3 &  (q - 1)/q^2 &  (q - 1)/q &  q - 1 &  q^2 - q  & q^3 -
    q^2\\
 1  & q - 1 &  q^2 - q &  2q^3 - q^2 &  -q^3 &  0 &  0\\
 1  & q - 1 &  q^2 - q &  -q^2 &  -q^3 &  0 &  0\\
 1  & q - 1 &  -q^3 + q^2 - q &  q^3 - q^2 &  q^4 - q^3 &  -q^4 &  0\\
 (-q^3 + 1)/q^3 &  (q - 1)/q^3 &  (q - 1)/q^2 &  (q - 1)/q &  q - 1 &  q^2 - q  & q^3  - q^2
\end{pmatrix}.
$$
It is straightforward to verify that $T$ is invertible and $T\cdot \sigma_{V}= \sigma_{W}\cdot T$. Thus 
$$\sigma_{V}=T^{-1}\cdot \sigma_{W}\cdot T.$$
Note that $T_1^{-1}=T^t$, thus for $1\leqslant i\leqslant 7$, the image $T_1^{-1}(x_{i-1})$ is equal to the dot product of $i$-th row of $T$ and the vector $(x_0,x_1,\dots,x_6)$. The images of  $T_1^{-1}$ (or $T_2^{-1}$) on the generators of $\C[W]^G$ above form a generating set of $10$ elements $g_1,g_2,\dots,g_{10}$ for $\C[V]^G$. More precisely,  
\begin{eqnarray*}
g_1&=&T_1^{-1}(x_{0})=x_0+  (q - 1)\cdot x_1+   (q^3 + q^2 - q)\cdot x_2+   (q^3 - q^2)\cdot x_3+   (q^4 - q^3)\cdot x_4   -q^4\cdot x_5\\
g_2&=&T_1^{-1}(x_{1})=x_0-x_1+q^3\cdot x_6\\
g_3&=&T_1^{-1}(x_{2})=\frac{q^3 + 1}{q^3}\cdot x_0+  \frac{q-1}{q^3}\cdot  x_1+\cdots\\
g_4&=&T_1^{-1}(x_{3})= x_0+ (q - 1)\cdot x_1 +  (q^2 - q)\cdot x_2+\cdots \\
g_5&=&T_2^{-1}(x_{4}^2)=[T_1^{-1}(x_{4})]^2=\left(x_0+  (q - 1)\cdot x_1+   (q^2 - q)\cdot x_2   -q^2\cdot x_3   -q^3\cdot x_4\right)^2\\
g_6&=&T_2^{-1}(x_{5}^2)=[T_1^{-1}(x_{5})]^2=\left(x_0+  (q - 1)\cdot x_1-  (q^3 - q^2 + q )\cdot x_2   +\cdots\right)^2\\
g_7&=&T_2^{-1}(x_{6}^2)=[T_1^{-1}(x_{6})]^2=\left(\frac{-q^3 + 1}{q^3} \cdot x_0+  \frac{q - 1}{q^3}\cdot x_1+\cdots\right)^2\\
g_8&=&T_2^{-1}(x_{4}x_5)=T_1^{-1}(x_{4})\cdot T_1^{-1}(x_{5})\\
&=&\left(x_0+  (q - 1)\cdot x_1+   (q^2 - q)\cdot x_2+\cdots\right)\left(x_0+  (q - 1)\cdot x_1-  (q^3 - q^2 + q )\cdot x_2   +\cdots\right)\\
g_9&=&T_2^{-1}(x_{4}x_6)=T_1^{-1}(x_{4})\cdot T_1^{-1}(x_{6})\\
&=&\left(x_0+  (q - 1)\cdot x_1+   (q^2 - q)\cdot x_2+\cdots\right)\left(\frac{-q^3 + 1}{q^3} \cdot x_0+  \frac{q - 1}{q^3}\cdot x_1+\cdots\right)\\
g_{10}&=&T_2^{-1}(x_{5}x_6)=T_1^{-1}(x_{5})\cdot T_1^{-1}(x_{6})\\
&=&\left(x_0+  (q - 1)\cdot x_1-  (q^3 - q^2 + q )\cdot x_2   +\cdots\right)\left(\frac{-q^3 + 1}{q^3} \cdot x_0+  \frac{q - 1}{q^3}\cdot x_1+\cdots\right).
\end{eqnarray*}
The invariant ring $\C[V]^G$ is also a complete intersection with three generating relations: $g_5\cdot g_6=g_8^2, 
g_5\cdot g_7=g_9^2$, and $g_6\cdot g_7=g_{10}^2$.
\hbo}\end{exam}

\begin{exam}\label{exam4.4}
{\rm
Let's consider the even self-dual codes in $M_{n,2}(\F_q)$; see  \cite[Section 2]{SA20} for more details.  The shape enumerator $H_{\CC}(x)$ of an even code $\CC$ in $M_{n,2}(\F_q)$ must be a polynomial satisfying
\begin{equation}
\label{ }
H_{\CC}(x)=H_{\CC}(x_0,x_1,x_2)=H_{\CC}(x_0,-x_1,x_2). 
\end{equation}
In other words, $H_{\CC}(x)$ is an invariant polynomial under the action of the group generated by
$$\tau:=\dia\{1,-1,1\};$$
see  \cite[Section 4.B]{SA20}. Recall that
$$
 \sigma=\frac{1}{q}\begin{pmatrix}
  1    & q-1  &q(q-1) \\
    1  & q-1 &-q\\
    1&-1&0
\end{pmatrix}
$$ and consider the group $G$ generated by $\tau$ and $\sigma$. Hence, the shape enumerator $H_{\CC}(x)$ of an even self-dual code $\CC$ in $M_{n,2}(\F_q)$ must be an invariant polynomial, under the action of $G$.

Note that $G$ is a finite group if $q=2$ and it is infinite, when $q\neq 2$. 

Now we consider the case $q=2$ and note that $|G|=6$. To find a generating set for the invariant ring $\C[x_0,x_1,x_2]^G$, 
we let $K$ be the subgroup of $G$ generated by $\uptau\cdot \upsigma$, where
$$\uptau\cdot \upsigma=\begin{pmatrix}
  1/2    & 1/2  &1 \\
    -1/2  & -1/2 &1\\
    1/2&-1/2&0
    \end{pmatrix}.$$
Note that $K$ is a normal subgroup of $G$ with order $3$. According to our experience, for many situations, computing invariants should become relatively easier when the group order decreases.  A Magma calculation \cite{BCP97} shows that
$\C[x_0,x_1,x_2]^K$ is generated by
\begin{eqnarray*}
f_1& := & x_0+x_2 \\
f_2 &:= &  x_0^2 + x_1^2 + 2x_2^2\\
f_3&:=&  x_0^3 + 3x_0x_2^2 + 3x_1^2x_2 + x_2^3,\\
h&:=&   x_0^2x_1 - 4x_0x_1x_2 - x_1^3 + 4x_1x_2^2
\end{eqnarray*}
subject to the unique relation
\begin{equation}
\label{ }
h^2+\frac{1}{3}\cdot f_1^6 - 2\cdot f_1^4f_2 + \frac{2}{3}\cdot f_1^3f_3 + 5\cdot f_1^2f_2^2 - 6\cdot f_1f_2f_3 - f_2^3 + 3\cdot f_3^2=0
\end{equation}
This relation also implies that $h^2\in \C[f_1,f_2,f_3].$ 

Note that the quotient group $G/K$ can be generated by $\uptau\cdot K$. As $f_1,f_2,f_3$ are fixed by $\uptau\cdot K$ and 
$h$ becomes $-h$ under the action of $\uptau\cdot K$, we have
$$\C[x_0,x_1,x_2]^G=(\C[x_0,x_1,x_2]^K)^{G/K}=\C[f_1,f_2,f_3,h]^{G/K}=\C[f_1,f_2,f_3,h^2].$$
Since $h^2$ can be expressed algebraically by $f_1,f_2,f_3$, we see that $\C[x_0,x_1,x_2]^G=\C[f_1,f_2,f_3]$.

As $\C[x_0,x_1,x_2]^G$ is a polynomial ring, a more theoretical method to prove that it can be generated by $\{f_1,f_2,f_3\}$ is to use a criterion appeared in \cite[Proposition 16]{Kem96}. A direct computation verifies that
the determinant of the Jacobian matrix of $\{f_1,f_2,f_3\}$ is nonzero, thus $f_1,f_2,f_3$ are algebraically independent over $\C$. 
Moreover, $|G|=6=\deg(f_1)\cdot \deg(f_2)\cdot\deg(f_3)$. Now applying \cite[Proposition 16]{Kem96} yields $\C[x_0,x_1,x_2]^G=\C[f_1,f_2,f_3]$.
\hbo}\end{exam}

\begin{rem}{\rm
In Example \ref{exam4.4}, the criterion \cite[Proposition 16]{Kem96} can be applied because the invariant ring is a polynomial algebra. If the invariant ring we are working on is not polynomial (actually, lot of invariant rings in the invariant theory of finite groups are not polynomial algebras), the "normal subgroup" technique will be more effective. 
\hbo}\end{rem}

\vspace{2mm}
\noindent \textbf{Acknowledgements}. 
This research was partially supported by the Algoma University under grant No. AURF-PT-40370-71. The authors would like to thank the anonymous referees and the editor for their careful reading, constructive comments, and suggestions. 
The first-named author would like thank Professor Simon Xu for his helpful conversations and encouragement.


\raggedright
\end{document}